\documentclass[sigconf]{acmart}

\AtBeginDocument{%
  \providecommand\BibTeX{{%
    \normalfont B\kern-0.5em{\scshape i\kern-0.25em b}\kern-0.8em\TeX}}}

\setcopyright{rightsretained}
\copyrightyear{}
\acmYear{}
\acmDOI{XXX}

\acmConference{}
\acmBooktitle{}
\acmPrice{}
\acmISBN{}

\begin{document}

\title{Nation-wide Mood: Large-scale Estimation of People's Mood from Web Search Query and Mobile Sensor Data}


\author{Wataru Sasaki}
\affiliation{%
  \institution{Keio Uniersity}
  \city{Fujisawa}
  \country{Japan}
}
\email{wataruew@sfc.keio.ac.jp}
\author{Hiroshi Kawane}
\affiliation{%
  \institution{Yahoo Japan Corporation}
  \city{Tokyo}
  \country{Japan}
}
\email{hkawane@yahoo-corp.jp}
\author{Satoko Miyahara}
\affiliation{%
  \institution{Yahoo Japan Corporation}
  \city{Tokyo}
  \country{Japan}
}
\email{smiyahar@yahoo-corp.jp}
\author{Kota Tsubouchi}
\affiliation{%
  \institution{Yahoo Japan Corporation}
  \city{Tokyo}
  \country{Japan}
}
\email{ktsubouc@yahoo-corp.jp}
\author{Tadashi Okoshi}
\affiliation{%
  \institution{Keio University}
  \city{Fujisawa}
  \country{Japan}
}
\email{slash@sfc.keio.ac.jp}

\renewcommand{\shortauthors}{Sasaki and Okoshi, et al.}

\begin{abstract}
The ability to estimate the current affective statuses of web users has considerable potential for the realization of user-centric services in the society. However, in real-world web services, it is difficult to determine the type of data to be used for such estimation, as well as collecting the ground truths of such affective statuses. We propose a novel method of such estimation based on the combined use of user web search queries and mobile sensor data. The system was deployed in our product server stack, and a large-scale data analysis with more than 11,000,000 users was conducted. 
Interestingly, our proposed ``Nation-wide Mood Score,'' which bundles the mood values of users across the country, 
(1) shows the daily and weekly rhythm of people's moods, 
(2) explains the ups and downs of people's moods in the COVID-19 pandemic, which is inversely synchronized to the number of new COVID-19 cases, and (3) detects the linkage with big news, which may affect many user's mood states simultaneously, even in a fine-grained time resolution, such as the order of hours.
\end{abstract}




\maketitle

\section{Introduction}
The ability for estimating the current affective state of web users is useful not only for realizing user-centric services tailored to specific statuses of individual users, but also for analyzing people's current affective states at a larger level, such as the entire web service, or people in different cities, regions, or demographic attributes. 
%
%
As an example, determining current people's positive (or negative) mood in web services at large, one can analyze relationships between people's such statuses and various types of societal events.

However, it is difficult to determine the affective statuses of web users outside a controlled in-lab configuration, particularly in the real-world situation of commercial web services. 
The first problem is {\bf the type of data from which the affective status of a user can be estimated}. Typically, sensing and determining the emotional state of a person require psycho-physiological data such as heart rate (HR)~\cite{mulder1992measurement}, heart rate variability (HRV), electrocardiogram (ECG), and electroencephalogram (EEG) data~\cite{ryu2005evaluation,wilson2002analysis}. 
However, the collection of such data in real-world conditions of mobile web users is not feasible owing to the low penetration rate of such sensors in society, the additional burden on users to use such devices, and lack of social acceptance for the collection of such data.  
The second problem is the {\bf difficulty associated with collecting the ground truth label on the user's affective status}. User annotation methodologies, including Experience Sampling Methodology (ESM) techniques, where users answer their subjective views on affective statuses, are widely used during the data-collection phase. 
However, this approach is not always effective because (1) the users may find it cumbersome to answer repeated questions, and (2) because the users may forget to answer the questionnaire. In particular, in commercial web services, it is not feasible to repeatedly send such questionnaires to users.

\begin{figure*}[t!]
\begin{center}
\includegraphics[width=0.85\linewidth]{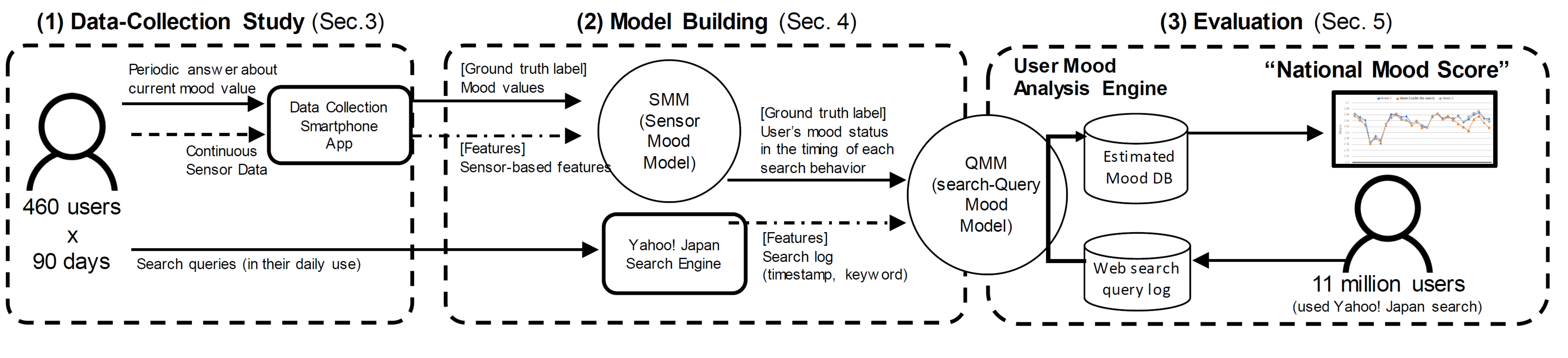}
\vspace{-0.4cm}
\caption{Framework of this research}
\vspace{-0.5cm}
\label{fig:researchframework}
\end{center}
\end{figure*}   

To solve these problems, in this paper, we show that the web users' affective status (concretely, mood) can be estimated in such a condition, based on {\bf a novel combined use of their web search queries and mobile sensor data}. 
To address the first problem, we pay special attention to users' query inputs to the web search engine as an easy-to-collect and noninvasive proxy feature to explain their mood states, focusing on the fact that almost all Internet users regularly use search engines in their daily lives. Because web services typically store the historical log of users' search queries on the server side, this methodology has an advantage in that it can be deployed immediately without having to wait for the widespread adoption of new types of sensors. 
To address the second problem, we used a novel two-step mood classification with different types of models, namely the ``Sensor Mood Model (SMM)'' and ``Search-Query Mood Model (QMM)'' to generate more mood status labels for the model and boost the overall classification performance. 

Figure~\ref{fig:researchframework} illustrates our research structure.
(1) First, we conducted a preliminary data-collection study with 460 participants for 90 days to collect their continuous sensor data from their smartphones; a periodic subjective evaluation of their mood was also performed as ground truth annotation (Section~\ref{sec:DataCollection}). 

(2) Next, we built our first model ``SMM,'' which estimates the participant's mood statuses from specific temporal frames in which both sensor data and the user annotation were successfully collected (Section~\ref{sec:ModelBuilding}).
With the built SMM, we can estimate each participant's mood status for all of the time frames during the data-collection study period, as well as the number of mood status labels. 
Then, by combining the web search logs of the 460 participants during the study period and mood status (both the users' original annotation and SMM's outputs), we trained the second model ``QMM,'' which estimates the mood of a user from their search query data.

(3) We built the analysis system in our server stack. This calculates mood scores of registered Yahoo! Japan's users (more than 11 million users) every 3 hours by processing their search queries with QMM. The average of all users' scores is calculated as the ``Nation-wide Mood Score.'' 
Using this, we evaluate the change in Nation-wide Mood Scores over time, and discuss how it matches the events and happenings in society. Interestingly, we found that (i) the score obtained based on our proposed algorithm shows the weekly rhythm of people's mood (which drops every 1st working day of the week and increases again every weekend) (Section~\ref{sec:result2}) and (ii) the longer-term trace of people's mood in the COVID-19 pandemic period in 2020 is inversely synchronized to the daily number of COVID-19 new cases (Section~\ref{sec:result3}). We also found that prefectures with more COVID-19 cases experienced deeper drop in their mood, according to our per-prefecture analysis.
Furthermore, (iii) we also found that the Nation-wide Mood Score can successfully capture the change in people’s moods influenced by significant news effect which simultaneously affects many users’ mood statuses (Section~\ref{sec:result4}).
{\bf To the best of our knowledge, our work is the first to build and deploy a machine-learning-based system for estimating people's affective mood status from the real-world data, along with evaluation on a large scale and in a long term.}


\vspace{-0.3cm}
\section{Related Work}
\label{sec:RelatedWork}

Extensive studies on emotion began to be conducted in the 19th century, 
with a well-known study conducted by Darwin~\cite{ekman2006darwin}, who reported that emotion is a product of evolution, and that emotions induce actions favorable to survival~\cite{darwin1998expression}.
Numerous emotional modalities and their respective physiological responses have been studied~\cite{royet2000emotional, craig2002you}.
Emotional states are known to affect cognitive and athletic abilities, and are reported to affect both human–human and human–machine interactions.

Picard generalized this research field as ``affective  computing''~\cite{picard1997affective}. 
Many studies and systems have been proposed to detect and utilize the emotions of users~\cite{chang2011s, de2003real, khan2013towards, likamwa2013moodscope, scheirer2002frustrating} in this field.
Several methods for determining user emotion have been proposed, which focus on physical characteristics~\cite{kwon2007emotion,Goncalves:2014:PTD:2632048.2636067}, and text data~\cite{Mohammad:2010:EEC:1860631.1860635,wang2012harnessing}.
In our research, we focus on the mood status of users. Mood is related, but is different from emotion in several ways~\cite{beedie2005distinctions}, including the tendency for mood to last longer than emotion, and that it is usually a cumulative reaction, while emotion is a more spontaneous reaction. 

Mood estimating research on smartphones have being conducted in the recent years.
The advantages of this type of approach are sensing capability of smartphones with a wide variety of embedded sensors, to be able to reduce the burden on users to wear various dedicated sensors, and easiness to use people's devices, thanks to the high penetration rate (e.g., more than 95\% in Japan)
Most studies have constructed a classification model that determines moods from the user's contextual data obtained from the smartphone sensor data and self-reporting annotation (mainly via Experience Sampling Method) by the user~\cite{ma2012daily}.
MoodScope~\cite{likamwa2013moodscope} investigated the effects of the user context on the mood of a user based on the smartphone sensor data.
In addition to emotion and mood, various types of internal statuses of the users, such as ``interruptibility''~\cite{pejovic2014interruptme,okoshi2019real}, have been recognized and estimated from the smartphone data.
Other types of sensing modality for emotion estimation is facial expressions in the image data. Such research has been widely conducted~\cite{fasel2003automatic}, mainly by using Facial Action Coding System~\cite{ekman1997face}. And some research on the smartphone~\cite{suk2015real} platform have been also performed recently.
In spite of such previous literature and systems that use smartphones with multiple types of sensors and modalities, deployment and evaluation of such methodologies in large-scale (e.g., with 10 million users) product web service in the real world have not been fully researched/explored, and that is where the novelty of this research lies.

Focusing on the data on the web, researchers are continuously working on estimating the emotional states of users by analyzing text from the users. 
There are researches to estimate user's affective status from their search queries on the web search engine, relying on emotion-related keywords (e.g., adjectives from which emotion has been easily inferred)~\cite{kozareva2007ua}. 
However, in the real world situation of the web search engine, it is difficult to conduct such estimation since most of the search queries are with only a few words (1.9 according to \cite{Tsubouchi2020}), and mainly with nouns. 
More recently, analysis of text data on social networks such as Twitter~\cite{wang2012harnessing, bollen2011modeling} and Facebook~\cite{kramer2012spread, kramer2014experimental} are actively conducted, since this type of text data is considered to contain more words and longer sentences possibly including words related their affective status. 
In addition, under the influence of COVID-19 pandemic, research on estimating depression mood from social media has been actively conducted in recent years~\cite{10.1007/978-3-030-65390-3_46, 9325873}.
However, users' data on such major social networks are not always accessible from the viewpoint of individual web services unless each user links their account to their social media account and permits the web service to access and analyze their social media update. 
Our approach, in contrast, relies on queries to the search function which is considered to be much more commonly deployed on each web site, thus is easier to be introduced by each web service. The novelty of our method lies also in the combined use of different types of mood estimation models from smartphone sensors and search queries.

\section{Data Collection Study}
\label{sec:DataCollection}
We first conducted a data-collection study with 460 users for 90 days (from October to December of 2019). We collected continuous data from various types of smartphone sensors as well as the user's subjective mood evaluation (up to 6 times a day) as the ground truth label. 

\vspace{-0.2cm}
\subsection{Participants}
For the study, participants were recruited through an external agency. 
The recruitment criteria were as follows: (1) the age should be in the range of 18-59 years,  (2) must own an active Yahoo! Japan registered account, (3) must have the ability to use the Yahoo! Japan search functionality once or more times per week and should have performed a search at least once in the last month, (4) must own and use an Apple iOS smartphone as a private primary phone in their daily life, and (5) must be the user of an iPhone 7 or later and iOS version 12 or above. During the recruitment process, the participants were informed that this study was ``an experiment about your condition.'' 

We successfully recruited 460 participants.
Finally, we used data from 338 users, excluding those who stopped data collection during the experiment or who could not collect enough data.
The participants are from different areas of Japan (geographically distributed),
consisted of 121 men and 217 women,
aged between 19 and 54 years (average:36.92), 
and with various types of occupations (civil servant:4.4\%, business executive:0.9\%, company employee (clerical:22.2\%, technical:8.0\%, other:16.6\%), self-employed:3.6\%, freelancer:0.9\%, homemaker:28.7\%, part-time job:9.2\%, student:1.8\%, unemployed:1.8\%, other:2.1\%). 



\begin{figure}[!t]
\begin{center}
\includegraphics[width=0.8\linewidth]{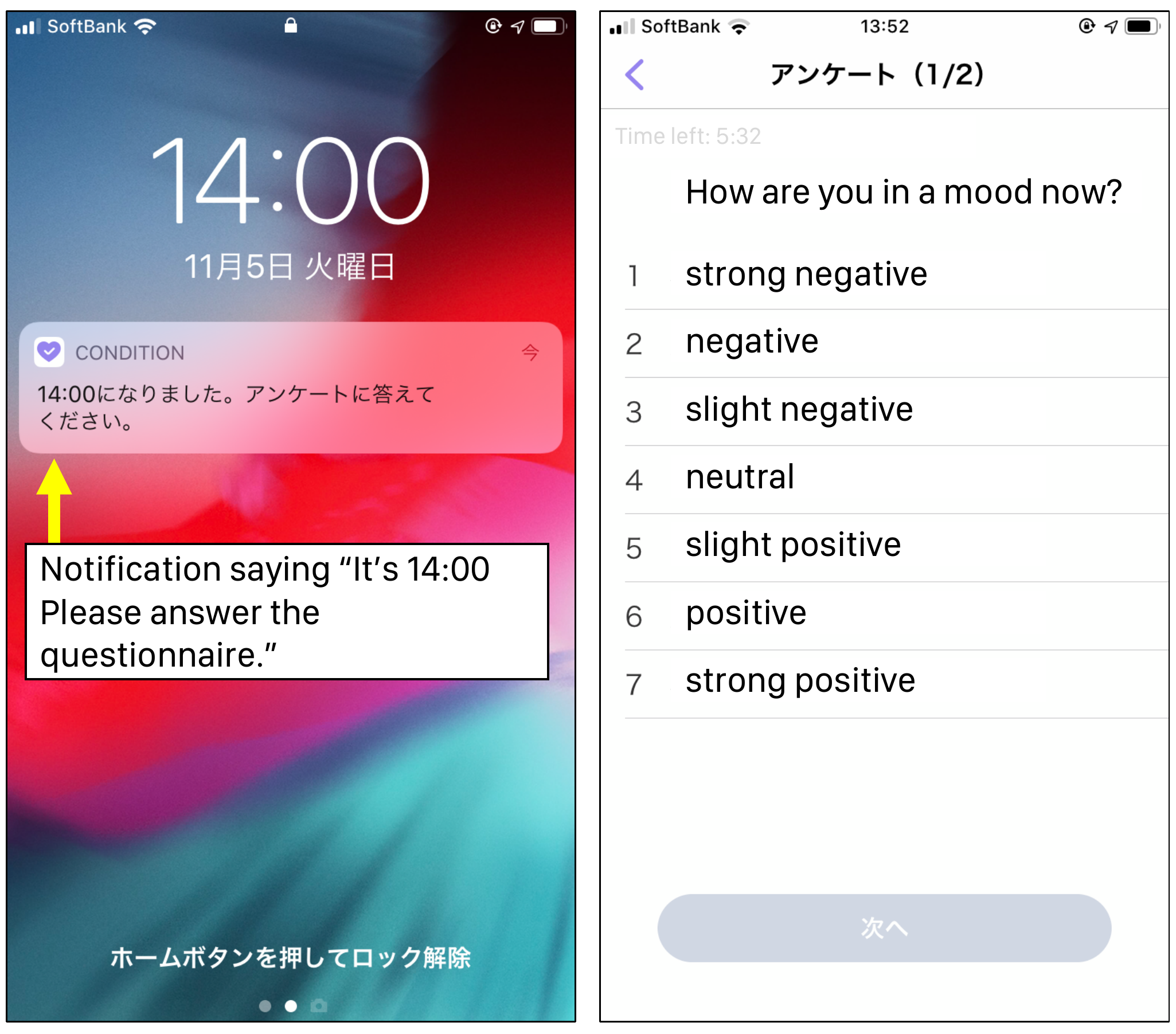}
\vspace{-0.2cm}
\caption{Screenshot of Data Collection Application}
\vspace{-0.5cm}
\label{fig:app_screenshot}
\end{center}
\end{figure}   

\vspace{-0.2cm}
\subsection{Experimental Setup}
In this study, we developed a dedicated smartphone application, as illustrated in Figure~\ref{fig:app_screenshot}. The application was developed for the iOS platform for several reasons. First, the market share of iOS is higher than that of Android in the Japanese market; thus, the recruitment of participants is easier. 
Second, the number of Apple iPhone models (e.g., iPhone 7, 7Plus, 8, 8Plus, X, and 11) is smaller than that of Android phones (hundreds of models by dozens of manufacturers with different OS-level optimization in power management, sensing, etc.). Thus, we can easily test the application with such phone models to achieve higher execution stability. 
Third, thanks to the iOS AWARE Framework~\cite{ferreira2015aware,nishiyama2020ios}, we can implement and deploy an application that can continuously collect various sensor data in spite of the fact that iOS is a rather strict environment as a sensing platform than Android.

Once the application is installed on the smartphone of a user, it continuously collects multiple types of data from the embedded sensors of the phone, as detailed in Table~\ref{session_data}, and periodically uploads the data to the server. 

The application can also issue a notification (as shown on the left-hand side of Figure~\ref{fig:app_screenshot}) at intervals configured by the developer to initiate the experience sampling method (ESM)-style data collection. (Note that the actual ``delivery'' timing of such notifications to the smartphone devices differ for each user owing to the real-world random behavior of the application on each device along with various types of network-related variables. This behavior generates some randomness to the timing in order to reduce possible bias in the experiment.) Once the user responds to the notification, the application opens a questionnaire on which the user can report his/her current mood status on a 7-level Likert scale (1. strongly negative, 2. negative, 3. moderately negative, 4. neutral, 5. moderately positive, 6. positive, and 7. strongly positive).

\begin{table}[t]
  \begin{center}
  \caption{Sensor Type and Sensing Frequency}
  \vspace{-0.3cm}
  \small 
    \begin{tabular}{cc} \toprule
Sensor Type & Sensing Frequency\\ \midrule
Accelerometer & 10Hz \\
Barometer & 1Hz \\
Battery status & 1Hz \\
Gyroscope & 10Hz \\
Location & 1/180Hz\\
Network type & (changed event-based)\\
Weather (from OpenWeather) & 1/60Hz\\
Screen status(On/Off) & (changed event-based)\\
\bottomrule
    \end{tabular}
    \label{session_data}
  \end{center}
\vspace{-0.5cm}
\end{table}


\subsection{Experimental Procedure}
Our experimental procedure consisted of the following three steps:  
(1) Each participant had a telephone meeting with a study researcher at the beginning of the study and received instructions including overview, purpose, type of sensor and query data collected, schedule and tasks about the study, which was followed by the signing of a consent form. Because the experiment involved the participants' sensitive information and data, including sensor data and subjective mood status, we were careful to protect the participants' ethical-related rights by carefully communicating with the participants. 

(2) Next, the participants were asked to install and launch our software on their smartphones. They were asked to grant the following permissions to the application: push notification, motion and fitness activity, and location (configured as ``always'') data sensing feature of the iOS platform.

(3) After the initial meeting, the 30-day study period started. 
During this period, a push notification appeared six times daily. (at approximately 8:00 AM, 10:00 AM, noon, 2:00 PM, 4:00 PM, and 6:00 PM, but the actual notification timings on the client side varies over several minutes.)
Each user was asked to proceed with the survey within 2 hours after the delivery of each notification. 
When the participant opened a notification, the application screen (Figure~\ref{fig:app_screenshot}) appeared, and the user was queried about their affective mood on a 7-level Likert scale. The participant selected the status, and after a confirmation prompt, the answer was submitted to the server.

\subsection{Reward}
We created an instant point reward system for the application. Each participant scored 0, 20, 30, or 40 points for 0-3, 4, 5, or 6 answers, respectively, on a daily basis, and the reward points accumulated throughout the study period. When a participant reached the configured minimum total reward points, that is, 1500 points (by answering four answers every day for 75 days or six answers every day for 38 days), they received a payment of 3000 yen. 
They received an additional payment of 2000 yen when they exceeded 2000 points.

\section{Model Building}
\label{sec:ModelBuilding}
After the data collection, this section describes our model building for two different models, the sensor mood model (SMM) and the search-query mood model (QMM).
As introduced in Figure 1, SMM classifies a user's mood from a set of features computed from sensor data. Using the built SMM model with more training data, we built a QMM that estimates the user's mood from search queries.

\subsection{Sensor Mood Model (SMM)} 
\label{sec:SMM}
To build SMM, we use an approach widely used in the activity recognition research area~\cite{bao2004activity}) to build a classifier with time-frame-based feature extraction from time-series sensor data. 

\subsubsection{Feature Extraction}
First, from the raw sensor data obtained from the data collection study, we extracted a set of features on a 3-hour time window. Table~\ref{tbl:features} summarizes the number of features extracted along with several representative feature types. The types of extracted features vary depending on the sensor type.

\begin{table}[t]
  \begin{center}
      \caption{Extracted Features}
\vspace{-0.35cm}
\resizebox{1.0\columnwidth}{!}{
    \begin{tabular}{ccl} \toprule
{\bf Sensor Type} & {\bf Number of Features} & {\bf Representative Features}\\ \midrule
Accelerometer, Gyroscope & 23 & 
\begin{tabular}{l}
(mean, std, median, min, max) magnitude \\ mean (each axis) \\ variance (each axis) \\ variance (each axis) \\ skew (each axis) \\ kurtosis(each axis) \\ correlation (xy, yz, zx) \\ covariance (xy, yz, zx) \\
\end{tabular}\\ \midrule
Barometer & 5 &
\begin{tabular}{l}(mean, std, median, min, max) magnitude \end{tabular}\\ \midrule
Battery status & 7 &
\begin{tabular}{l}
(mean, std, median, min, max) battery level \\
number of charge times \\
duration of charge minutes \\
\end{tabular}\\ \midrule
Location & 12 & 
\begin{tabular}{l}
location entropy \\
number of location transitions\\
moving time percent\\
\end{tabular}\\ \midrule
Network type & 5 &
\begin{tabular}{l}
number of WiFi connectivity established\\
number of mobile connectivity established\\
most frequent network type \\
rate of WiFi \\
rate of mobile network\\
\end{tabular}\\ \midrule
\begin{tabular}{c}
Weather\\ (from OpenWeather) 
\end{tabular}
& 50 &
\begin{tabular}{l}
weather type \\
(mean, std, median, min, max) temperature \\
(mean, std, median, min, max) humidity \\
\end{tabular}\\ \midrule
Screen status(On/Off) & 11 & 
\begin{tabular}{l}
number of unlocks (per minute) \\
number of interaction (per minute)\\
\end{tabular}\\ \bottomrule
    \end{tabular}}
    \label{tbl:features}
  \end{center}
\vspace{-0.3cm}
\end{table}

\subsubsection{Model Building}
Then, we constructed a supervised machine-learning model.
For each user, for each 3-hour frame, a pair of the calculated features and a user's self-reported mood status (as the ground truth label, as long as it is available) are combined and input to the ML algorithm as training data. 
We aimed to build a single SMM from all available training data obtained from all the users. 

In this model building, we treat the self-reported mood status as a three-class classification problem in consideration of practical operability.
The collected mood answers (originally on the 7-level Likert scale) were assigned to three different labels, $-1$ for ``strongly negative,'' ``negative,'' and ``moderately negative,'' $0$ for ``neutral,'' and $+1$ for ``strongly positive,'' ``positive'' and ``moderately positive.''
We chose Random Forest~\cite{RandomForest} for the machine-learning algorithm, which revealed the best classification performance compared with others.
To improve the model performance, we optimized the hyperparameters such as ``criterion'', ``bootstrap'', ``max\_features'', ``max\_depth'', ``max\_leaf\_nodes'', ``n\_estimators'', ``min\_samples\_split'', and ``min\_samples\_leaf'' using Optuna~\cite{akiba2019optuna} which is a software framework for automating hyperparameter optimization.

\subsubsection{Model Performance}
In order to evaluate the SMM's performance, we conducted a 5-fold cross-validation.
The data were randomly divided into 80\% training data set and 20\% test data set.
The model achieved an accuracy of 72.0\%.
Table~\ref{tbl:mood_perfomance} shows the detailed results of the overall performance evaluation.
The imbalanced results depending on the label were due to the different number of collected data. 
Therefore, considering the imbalanced result, we calculated the macro-F1.
The macro-F1 score was 0.60.

We also calculated the feature importance of the SMM. 
The top-5 feature types and importance scores were ``mean-x-gyro''(0.053), ``mean-z-gyro'' (0.052),  ``mean-y-gyro'' (0.051), ``median-magnitude-acc'' (0.037) and ``median-magnitude-gyro'' (0.025). 
We confirmed that there were many features with high importance score were related to gyroscope sensor.

\begin{table}[t]
  \begin{center}
      \caption{Performance of Sensor Mood Model (SMM)}
    \vspace{-0.3cm}
\resizebox{0.65\columnwidth}{!}{
    \begin{tabular}{cccc} \toprule
    Label & Precision & Recall & F1-score \\ \midrule
    -1 & 0.42 & 0.35 & 0.38 \\ 
    0 & 0.59 & 0.62 & 0.61 \\
    1 & 0.82 & 0.83 & 0.82 \\ \midrule
    micro avg & 0.73 & 0.72 & 0.72 \\ 
    macro avg & 0.61 & 0.60 & 0.60 \\ \bottomrule
    \end{tabular}
    }
    \label{tbl:mood_perfomance}
  \end{center}
  \vspace{-0.3cm}
\end{table}


\subsection{Query Mood Model (QMM)} 

We define the mood score that represents the score of the user's mood status.
After we built the SMM model (in Section~\ref{sec:SMM}), we have two different types of mood scores, namely (a) the scores answered by the data collection participants with the questionnaire, and (b) the scores estimated by SMM based on the collected sensor data.
QMM is a model that examines the relationship between a user's web search query and the user's  mood score during the search behavior. 
After training, the user's mood score is classified from their search query data.
In this section, we explain the concept of QMM and explain how the additional use of (b) increases the QMM performance.

\subsubsection{Model Building}
For each fixed time range (here, we refer to it as a ``session''), QMM is trained from the data of a user's mood score and search behavior during the session. 

To investigate the validity of the trained QMMs qualitatively, we employed as a model logistic regression, which is a typical example of a ``white box'' model with high model interpretability.
Note that it is not necessary to specify the logistic regression as a training scheme in the actual operation; we believe that non-linear SVMs and decision tree-based regressions (such as Xgboost, which are specialized for performance) are also effective.

One session was defined as one record, and training was performed using the following regression equation: 

\begin{eqnarray}
    y = \theta_0 + \theta_1 x_1 + \theta_2 x_2 + \dots + \theta_n x_n ,
\end{eqnarray}
where $y$ is the mood score and $\theta_k$ is the learned weight. $x_k$ is the search query assigned to a feature, and it has a value of 1 if it was searched in that session, and 0 if it was not.
$x_k$ indicates only whether the query is searched, and does not indicate the number of searches.
For multi search words, we extracted each word and counted one for each according to the rule of $x_k$.

\subsubsection{Combination with SMM}
During the model training, some sessions miss the ``mood score,'' mainly because the user's raw answer to the mood questionnaire is not available, although both ``search behavior'' and ``mood score'' are needed as training data. 
In such a situation, the use of pre-trained SMM along with collected sensor data, is an effective means of generating mood scores to be used for the training.
In contrast to the mood scores from the users with the questionnaire, the mood score estimated by SMM can be prepared constantly since their smartphone sensor data were continuously collected by our application throughout the entire study period (except for some irregular cases where our application was not working). 
Therefore, the SMM can be used to supplement missing mood scores, essentially creating labels for all periods of time during the data collection study, and finally, increasing the amount of training data for QMM.

We set the length of a session to 3 hours. All search queries retrieved within the 3-hour period were used as features. Because SMM-based mood scores are available 24 h a day, the length of each session can be shortened (fine-grained). 
However, if the length is too short, there is a risk that the questionnaire-based mood scores of the comparative method (QMM to be trained without the output from SMM; it is to be used in our comparative evaluation in Section~\ref{sec:EvaluationNation}) may become too sparse to be learned.
Hence, the sessions needed to be reasonably long. 
From this discussion, we decided to use a session length of 3 hour in this study.

\subsubsection{Cross-validation Performance}
We built two QMM models, one trained only from the questionnaire answer data, and the other with additional training data based on the SMM outputs. 
For both models, the condition of the performance evaluation is as follows. 
The data were randomly divided into 80\% training data and 20\% evaluation data. 
Considering the effect of randomness, the evaluations were conducted for 10 times. 
The training data were balanced so that the amount of positive and negative data was the same before training. 
For the evaluation data, we did not perform balancing.

The results are shown in Table~\ref{tab:acc}, which clearly confirms the effectiveness of the SMM. 
Compared to our baseline QMM without SMM use, the accuracy increases from 87.0\% to 94.2\% in the case with additional data brought by SMM. 
The table also shows that the amount of training data has been more than doubled by SMM, indicating that the more than doubling of the dataset used for training by SMM contributed greatly to the significant improvement in prediction accuracy.
From these results, we decide to adopt a QMM trained with SMM in this research, and proceed to our evaluation experiments. 

\begin{table}[t]
\caption{Performance of search-Query Mood Model (QMM)}
\begin{center}
\vspace{-0.3cm}
\resizebox{0.7\columnwidth}{!}{
\begin{tabular}{lcc}
\toprule
\multicolumn{1}{c}{} & \# of data & accuracy \\ \midrule
QMM (with SMM (proposed))  & {\bf 52,252}     & {\bf 94.2\%}     \\
QMM (without SMM)          & 27,402     & 87.0\%     \\ \bottomrule
\end{tabular}}
\end{center}
\label{tab:acc}
\vspace{-0.3cm}
\end{table}

\subsubsection{Query Features Constructed for the QMM}
When we investigate inside the trained QMM, we can qualitatively confirm that the model actually represents the mood.
We collected approximately 81,000 unique search queries in the experiment.
The number of queries weighted in the trained model (out of all 81,000 query words) was 217.
Table~\ref{tab:qmm_example} shows example of representative queries that had high weights in the trained QMM. 
(Note that all descriptions in this table are translated to English; the original queries are in Japanese. The record with \$ is the explanation of the query content, rather than the actual content itself, and this is to protect the actual business and web site.) 
For example, the top positive words include the names of shopping (price) comparison web sites for home appliances and restaurants, the name of a smartphone game for walking around outside, indicating the importance of users to actively obtain a variety of information. On the other hand, direct expressions such as ``pachinko'' (Japanese-style pinball gambling) and ``I want to die'' were ranked as negative words, not to mention that they fit our intuition.

\begin{table}[t]
\caption{Query Features with High Weight}
\begin{center}
\vspace{-0.3cm}
\resizebox{0.9\columnwidth}{!}{
\begin{tabular}{cc}
\toprule
positive queries & negative queries \\ \midrule
\$shopping comparison web site\$   &  ``I want to die''     \\
``baby''     &  ``pachinko'' (Japanese pinball gamble) \\
\$Smartphone walk-around game\$     &    ``credit card'' \\
\$celebrity blog web site\$     &    ``temporary worker'' \\
\$gourmet comparison web site\$ & ``headache'' \\
\$web site for checking utility usage/price\$ & \$fraudulent site\$ \\
  \bottomrule
\end{tabular}}
\vspace{-0.3cm}
\end{center}
\label{tab:qmm_example}
\end{table}

\section{Evaluation: Nation-wide Mood}
\label{sec:EvaluationNation}

\subsection{Overview}
To examine the value of calculating the mood scores of a large number of registered users, we assessed changes in mood scores across several different timescales and how they relate to social phenomena.
All data analysis was conducted on Yahoo! Japan's internal servers after anonymization 
so that only the statistical information could be viewed by the research members.

\subsubsection{Nation-wide Mood Score}
The index score we use for this evaluation is ``Nation-wide Mood Score,'' which is the average mood score of all target users (approximately 11,000,000 users all over Japan). 
Since we use users' historical search query logs, the population from which the score is calculated comprises Yahoo! Japan's registered users who used our search engine at least once, for each particular day or time. (Thus, the exact number of target users changes continuously.)
Note that the mood score as an calculated output from trained QMM and user's search behavior can take a value without an upper or lower limit theoretically because the mood score is not sigmoid-fitted.

As a reasonable approximation, according to Yahoo! Japan's corporate data (opened to public, for the purpose of advertisement sales), demographics of the entire Yahoo! Japan's users are covering geographically all prefectures of Japan, with a mixture of male (48\%) and female (52\%), and covering the all age groups (male: 10s:2\%, 20s:16\%, 30s:17\%, 40s:22\%, 50s:18\%, 60s and above:25\%, female: 10s:1\%, 20s:16\%, 30s:17\%, 40s:22\%, 50s:18\%, 60s and above:25\%).

From such fact, by calculating and averaging the mood scores of all those users on a given day (or hour), we can derive a mood value that refer to as ``Nation-wide Mood Score'' in Japan on a daily (or hourly) basis.



\subsubsection{Comparative Method}
As described earlier, one of the challenges in this research is to show the effectiveness of the SMM output in the training of QMM. Thus, in this evaluation, we compare (1) the QMM trained with the SMM output as our proposed method, and (2) QMM without SMM as a comparative method.
Note that all the conditions (algorithm, hyper parameters, and split ratio between the data) are the same between these two methods.

\begin{figure}[t]
\begin{center}
\includegraphics[width=0.9\linewidth]{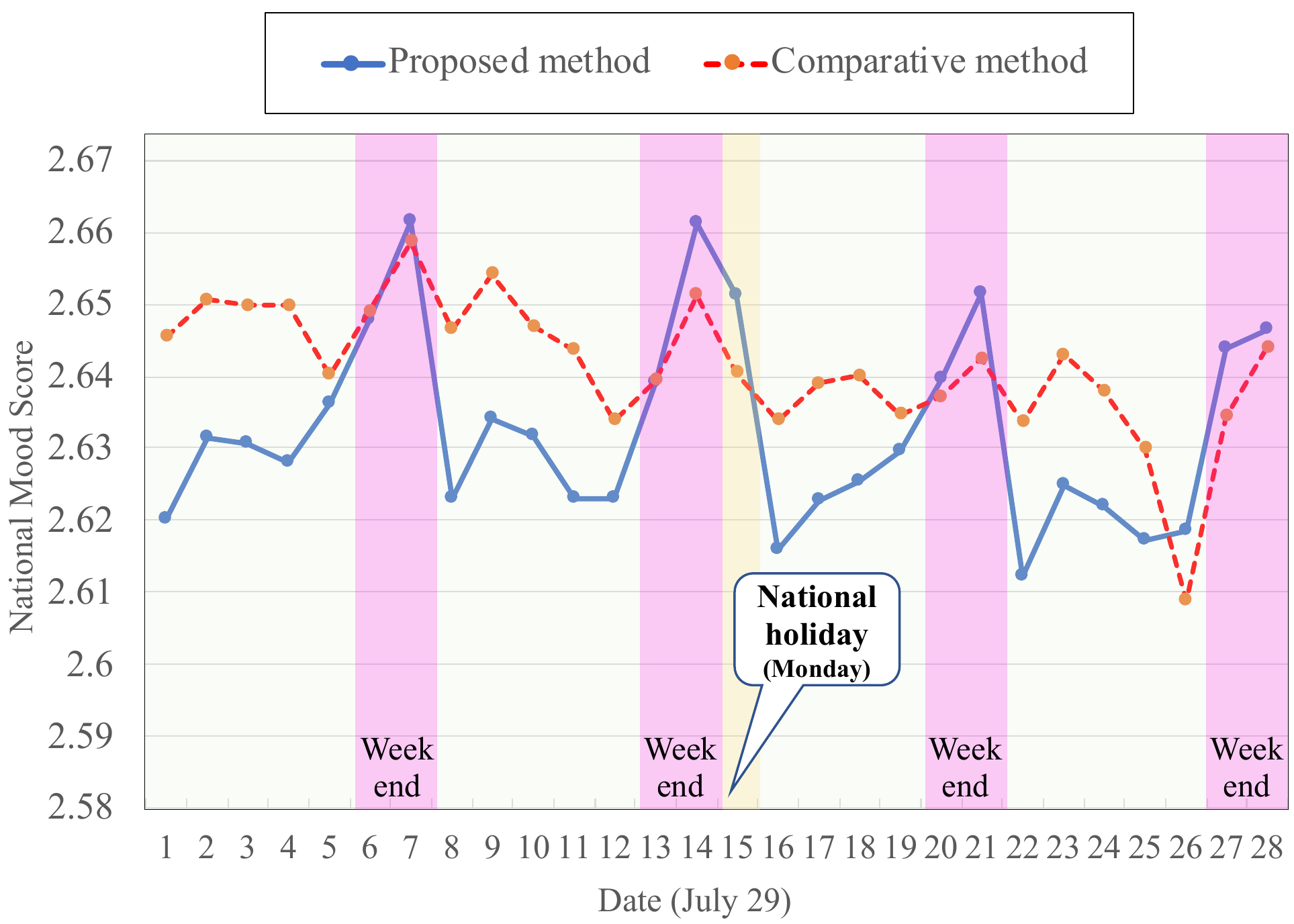}
\vspace{-0.3cm}
\caption{Weekly Rhythm of Daily Nation-wide Mood Score}
\label{fig:res2}
\vspace{-0.3cm}
\end{center}
\end{figure}



\subsection{Result 1: Weekly and Daily Mood Rhythms}
\label{sec:result2}
The first case is an analysis of the daily mood score trend for four weeks. We want to see how the nation-wide mood score changes within a month, which is a relatively short term. 
For this analysis, we used a dataset for the period from July 1, 2019 to July 28, 2019. 
Among the log data stored in our server, we carefully chose this period in order to 
exclude a period with a major breaking news story (such as a shocking big incident) that possibly influences the emotional status of many users.

\begin{figure*}[t]
\begin{center}
\includegraphics[width=0.9\linewidth]{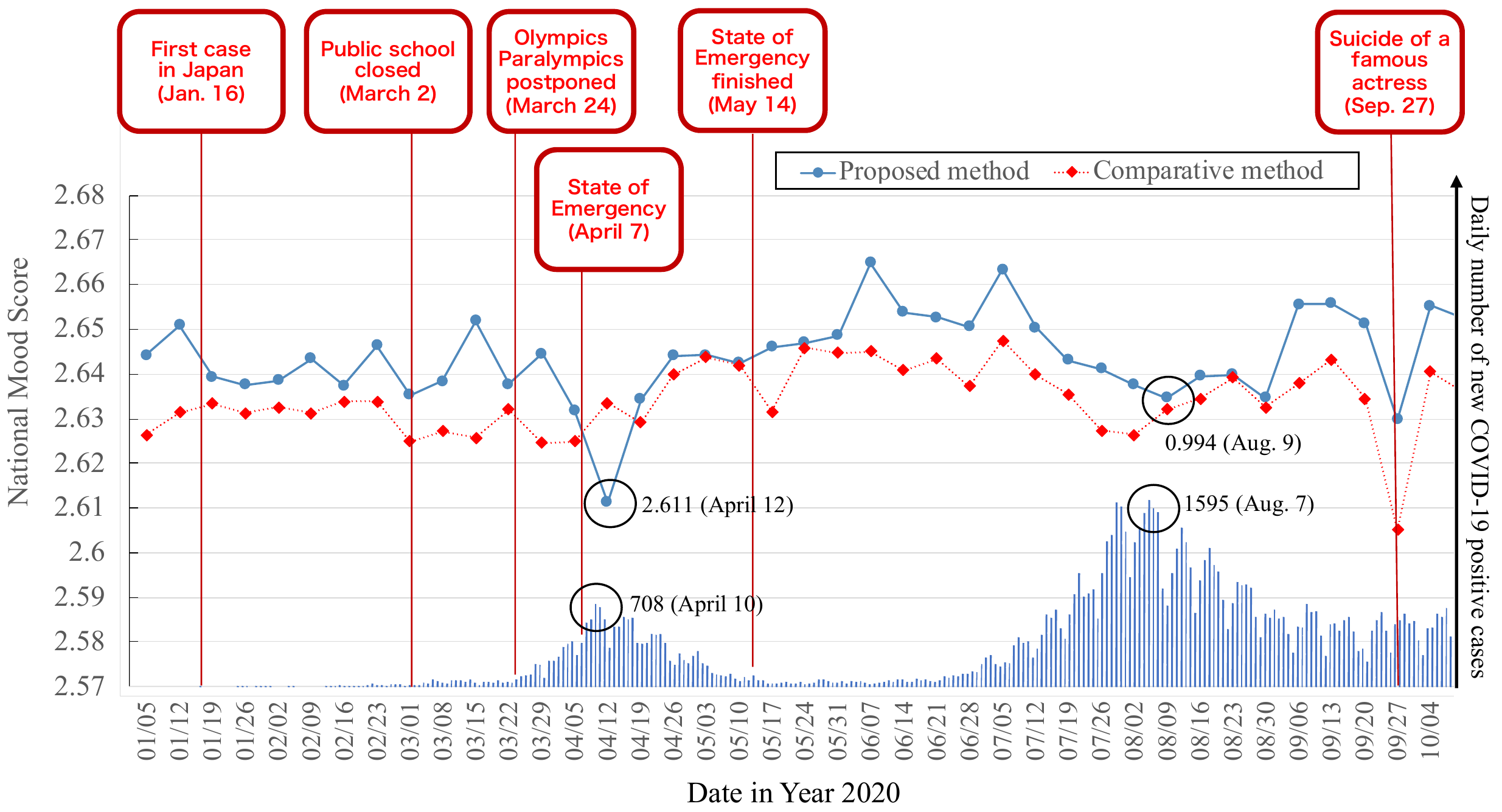}
\vspace{-0.4cm}
\caption{Relative Nation-wide Mood Score and Number of COVID-19 Patients in Year 2020}
\label{fig:res3}
\end{center}
\end{figure*}

Figure~\ref{fig:res2} shows the resulting nation-wide mood scores for this period. 
The x-axis represents the date while the y-axis represents the daily score.
A very interesting result in the figure is that, with our proposed method, the scores tend to be clearly positive on weekends, and more negative on Mondays when the workday begins (or Tuesdays when Monday is a holiday on July 15).


\begin{table}[h]
\begin{center}
\caption{Statistics on Weekly Rhythm}
\vspace{-0.4cm}
\label{tbl:weekstat}
\includegraphics[width=0.9\linewidth]{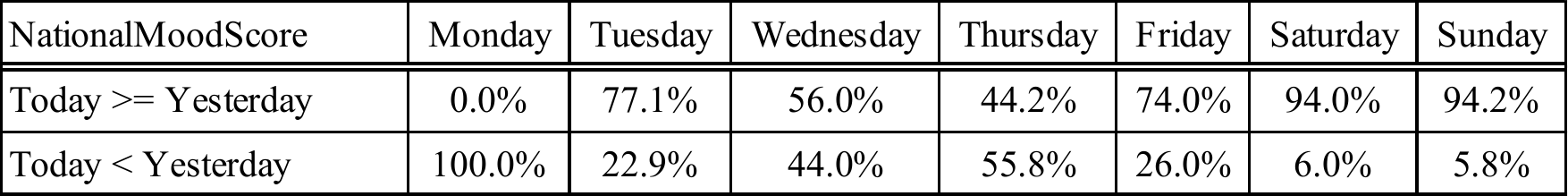}
\end{center}
\end{table}

Table~\ref{tbl:weekstat} shows the statistics on all days throughout this year. We omitted the national holidays, and counted the number of days whose daily score equals or is higher than the previous day's score. 
Surprisingly, on all (46 out of 46) Mondays, the score had decreased from the previous Sunday. Then, getting close to the weekend, especially on Fridays, Saturdays and Sundays, the scores got increased from the corresponding previous day.


\subsubsection{Discussion}
Although there exist no ground-truth data on a nation-wide mood, this tendency to feel better later in the week and then worsen again on Monday is considered to be an intuitive result in a society where many people work Monday through Friday, as there is the term ``Blue Monday.'' 
In fact, according to a white paper released by the Ministry of Health, Labour, and Welfare~\cite{JisatsuWP2019}, Monday is the day with the highest number of suicides in Japan. 
Another survey of 400 men and women found that the highest number of respondents in all age groups said they felt most depressed on Mondays~\cite{Grico}.

The proposed method looks to successfully express the rhythm of the mood change over the weekdays and weekends in an instructive manner, while such rhythm is not very clear in the comparative method. 
From this result, we conclude that the proposed method better explains the weekly mood rhythm than the comparative method.



\subsection{Result 2: COVID-19 and People's Mood}
\label{sec:result3}
Our second evaluation aims to reveal how the mood score have changed in the COVID-19 pandemic situation in 2020.


\subsubsection{Pandemic Waves and Nation-wide Mood}
Firstly we look into the waves of the pandemic and mood score in the nation-wide granularity. 
We chose mood score of Sundays because every Sunday is a holiday; otherwise, it was assumed that it would be difficult to discuss the analysis results owing to occasional holidays. 

Figure~\ref{fig:res3} shows the the daily number of COVID-19 new cases and the change of the Nation-wide Mood Score. The value in the y-axis (left side, lines) indicates the Nation-wide Mood Scores calculated by the proposed and comparative methods. 
Also, the bars show the daily number of new COVID-19 cases in Japan~\cite{JapanCOVID19OpenData}.

Surprisingly, the results show that for the proposed method, peaks of the waves of COVID-19 infection spread and degradation of the Nation-wide Mood Score are synchronized. 
The peak of the first COVID-19 wave was April 10 (the number of new cases: 708). 
On April 12, the mood score decreased to the first trough of 2.611. 
The peak of the second wave was August 7 (the number of new cases: 1,595). 
Only 2 days later, on August 9, the second trough of the Nation-wide Mood Score recorded 0.994.
However, we cannot confirm such a clear tendency with the comparative method. 

\begin{figure*}[h]
\begin{center}
\vspace{-0.1cm}
\includegraphics[width=0.9\linewidth]{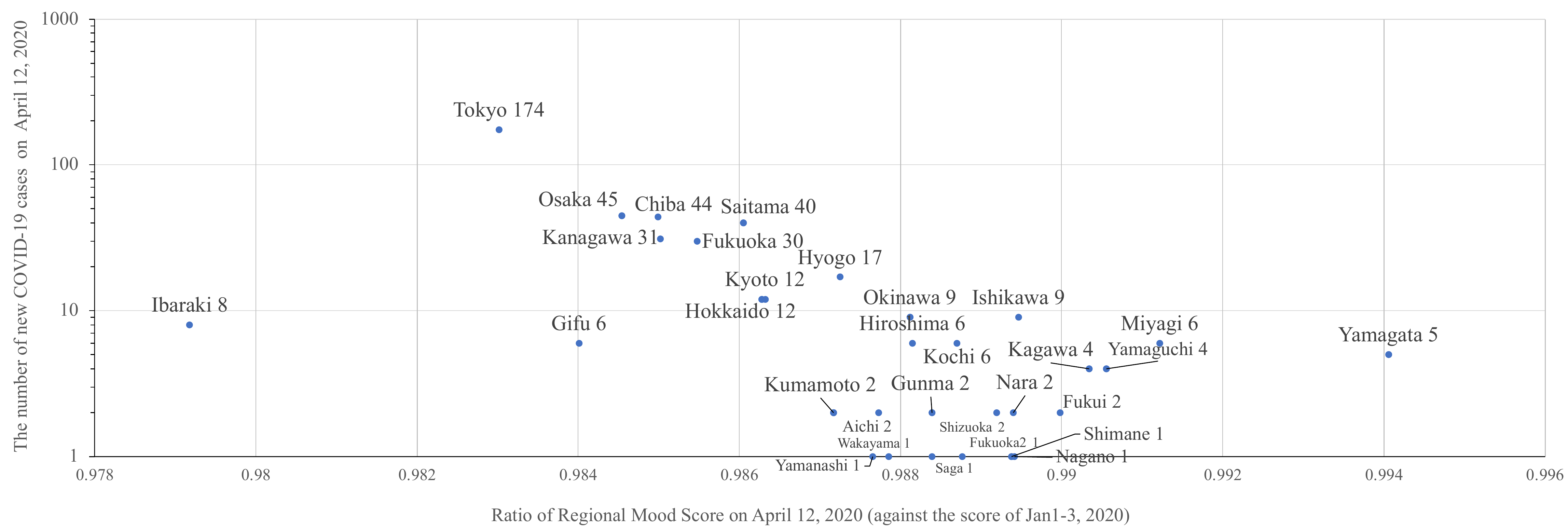}
\vspace{-0.3cm}
\caption{COVID-19 Cases at The Peak of The 1st Pandemic Wave and Drop in People's Mood in Each Prefecture}
\label{fig:region}
\vspace{-0.4cm}
\end{center}
\end{figure*}

It is very interesting to read the tendency for the mood to become more negative as the number of COVID-19 new cases increases, and vice versa.
In 2020, Japan originally planned to host the Tokyo Olympics and Paralympics. The mood in Japan was positive at the beginning of the year. However, COVID-19 then began to spread, the Olympics were postponed, and various economic activities were restricted.
The number of COVID-19 cases began to rise in Japan, and the public became frightened.
Then, as the first wave subsided, the mood was again positive for the restoration of economic activity.
However, when the second wave began, the mood became negative again.
Based on our analyses, this graph has successfully tracked the tumultuous changes in Japanese peoples’ moods in 2020.

\subsubsection{Differences in Mood Drops in Different Regions}
Next, we decided to examine the relationship between the drop in the mood score and the spread of the pandemic at a geographically finer level. 
During the the first wave of the pandemic (to April 12, 2020), most of the cases were found in big cities with the major airports with international flight routes, such as Tokyo and Osaka, along with their surrounding areas, such as Kanagawa, Chiba, and Saitama. On the other hand, other relatively rural areas away from these metropolitan areas generated no or very few positive cases. This difference might have affected the change of people's mood. We hypothesized that when mood scores were calculated by prefecture, the more COVID-19 cases a prefecture had, the more severe the drop in mood toward the peak of the first wave. Thus, we divided the same data from about 11 million users into each of 47 prefectures in Japan, according to each user's residential address address data (used for the payment of several Yahoo! Japan paid services) in their account profile, then calculated the daily ``Regional'' Mood Scores for each prefecture. 

Figure~\ref{fig:region} shows the result of our analysis. The y-axis indicates the number of new COVID-19 cases on April 12, 2020 (the peak Sunday of the 1st wave). Meanwhile, the x-axis shows the relative value of the Mood Scores in each prefecture on that day, compared with the average scores of the same region during the new year holidays (January 1 to 3, 2020). 
The lower this number (to the left) is, the more severe drop in mood toward the peak of the 1st wave. 
Very interestingly, a distribution from lower right to upper left is observed, indicating that the more COVID-19 positive prefectures have, the more severe the decline in mood, just as we hypothesized. We conducted the correlation analysis and found the Pearson correlation coefficient -0.43, indicating a moderate negative correlation. 

\subsubsection{Discussion}
Regarding these evaluations, the most remarkable point on our model (with the proposed model) is that {\bf the period of time when we collected sensor data and search queries for the model building was before the COVID-19 pandemic} (from October to December of 2019) as mentioned in Section~\ref{sec:DataCollection}.
This means that there is no possibility that search queries such as ``COVID-19'' could have been include in the trained QMM as features for negative mood values.
Again, there is no ground truth on the ``nation-wide mood.'' However, based on the facts that we can nevertheless observe the score trends (1) inversely synchronized with the number of COVID-19 cases and (2) negatively correlated to the COVID-19 cases, we conclude that the nation-wide mood score (with the proposed method) matches our intuition more.

\vspace{-0.15cm}
\subsection{Result 3: Big News That Affects Mood of Many Users At Once}
\label{sec:result4}

Interestingly, while we were analyzing the Nation-wide Mood Scores shown in Figure~\ref{fig:res3}, we also found that the Nation-wide Mood Score can successfully capture the change in people's mood influenced by a big news that affected many users’ mood at once. 

When we look at Figure~\ref{fig:res3}, we observe a sharp drop in the score on September 27, 2020. 
After some investigation, we realized that it was the day when the suicide of a famous Japanese 
actress was reported. From our investigation, at 8:29 AM on that day, the first tweet on Twitter~\cite{FirstTweetOfTakeuchiNews} reported that a breaking news text appeared on the TV program. (In Japan, big news stories are reported on the TV broadcast in the form of overlay texts on the TV screen.) Almost simultaneously, at 8:30 AM, the earliest breaking news article on her suicide was published on a web news site~\cite{FirstArticleOfTakeuchiNews}.

\begin{figure}[t]
\begin{center}
\vspace{-0.1cm}
\includegraphics[width=1.0\linewidth]{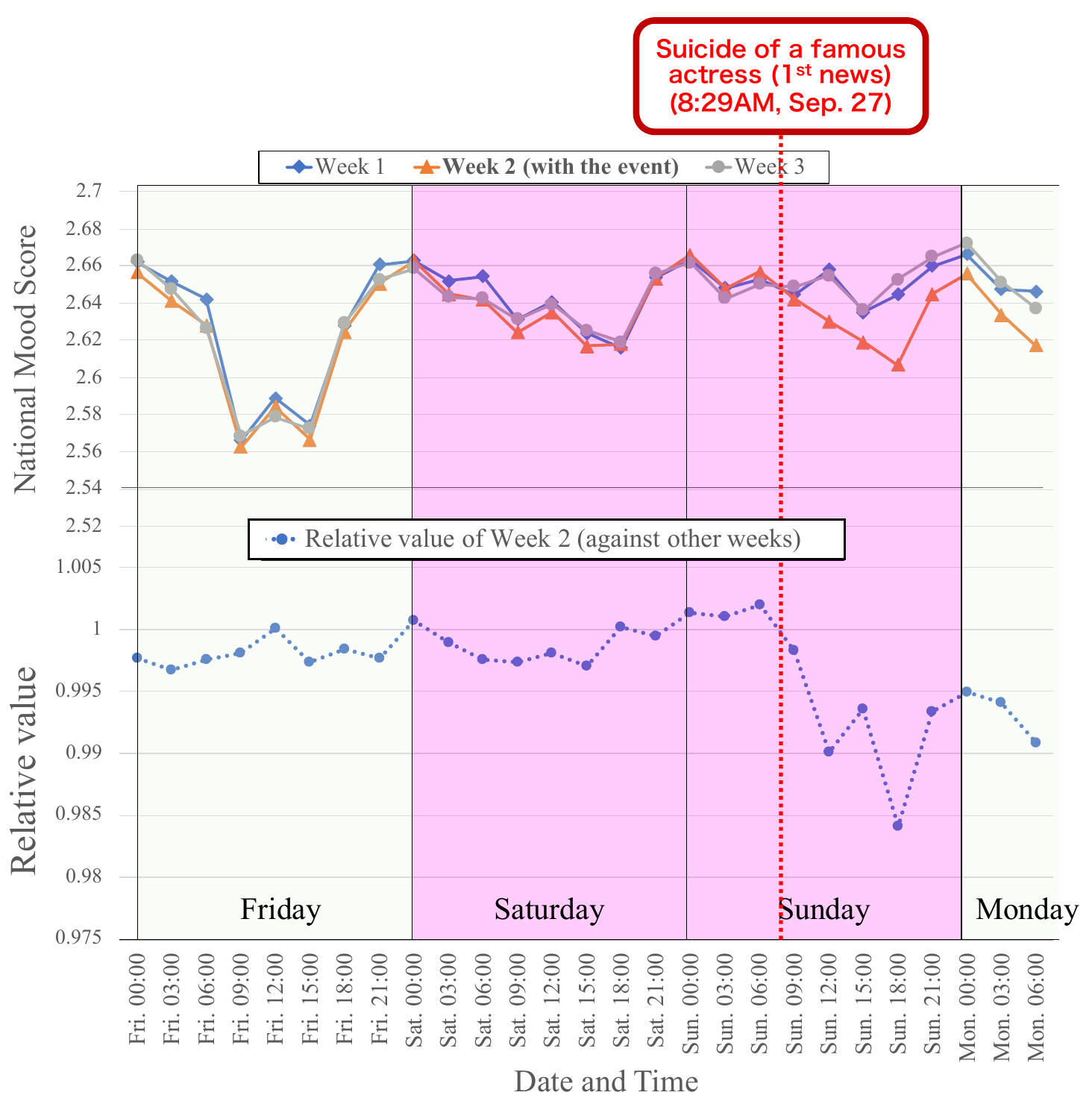}
\vspace{-0.7cm}
\caption{Score Gap on The Weekend of An Actress's Death}
\vspace{-0.7cm}
\label{fig:Takeuchi_gap}
\end{center}
\end{figure}


We observe the trace of the Nation-wide Mood Score on that day by comparing it with similar traces on other weekends. Figure~\ref{fig:Takeuchi_gap} (top) overlays the traces of the Nation-wide Mood Score for three different weekends, Week 1 (previous weekend), Week 2 (the target weekend), and Week 3 (the next weekend). When we look at the dotted line (which illustrates 8:30 AM of Sunday), the score of Week 2 clearly starts to decrease compared with the other two weeks. 
Figure~\ref{fig:Takeuchi_gap} (bottom) shows a trace of the relative score (the score of Week 2 divided by the average of the Week 1 and Week 3 values).
We can clearly observe that the mood becomes negative after 8:30 AM on September 27, 2020, when the news was reported.
From these results, we can conclude that the Nation-wide Mood Score is able to detect the linkage with a large event, which may affect many users' mood states, even in a fine-grained time resolution, such as the order of hours.

\section{Limitation and Future Work}
\label{sec:fw}
In this study, we created a model to estimate mood from web search queries, 
thereby analyzing the changes in weekly rhythm, COVID-19 pandemic, and big news. 
One of our future work is classification other affective statuses (beyond mood).
We expect that the same framework can be used to model a wide range of affective statuses. 
Another task is to improve the model performance. 
In this study, we adopted a white-box model to examine the effectiveness of the model qualitatively. We hope to employ models focused on precision and recall performance towards further performance improvement. 

We are interested to see how use of sensor and query data in the different time period affects the model performance of SMM and QMM. The relationship between user's mood, data obtained from the sensors, and query contents can be changed over time. 
In addition, there remains discussion as to whether an estimation model constructed from 460 participants' data could be applied to 11 million samples, even though the demographics of the 460 people and the 11 million user sample are not significantly different.
Increasing the number of user samples of the data that builds our model is one of the future works.

Importantly, the realization of real ``mood-aware web service'' needs to be designed very carefully, to protect user's privacy and to complying the ethics guideline in future research and service delivery. In any cases, various types of systematic design to protect users, such as advanced consent, opt-in strategy, transparency of the logic, will be necessary. 


\section{Conclusion}
\label{sec:Conclusion}
Affection-awareness is one of the key components of human-centric information services. However, in particular, in the real-world web field, estimating such statuses of the user is yet to be realized. We proposed a novel method of estimating web users’ moods based on the combined use of search queries and mobile sensor data. Our extensive data analysis revealed multiple interesting results, including the changes of Nation-wide Mood Scores in the weekly rhythm, in the COVID-19 pandemic situation, and in case of a big breaking news. 

\bibliographystyle{ACM-Reference-Format}
\bibliography{YahooEmo_WWW2022}

\end{document}